\documentclass[prd, aps, superscriptaddress, preprintnumbers, twocolumn, floatfix, nofootinbib]{revtex4}

\usepackage{amsfonts}
\usepackage{amsmath}
\usepackage{amssymb}
\usepackage{bm}
\usepackage{dcolumn}
\usepackage{graphicx}   
\usepackage[latin1]{inputenc}
\usepackage{latexsym}
\usepackage{rotating}
\usepackage{hyperref}
\usepackage{graphicx}
\usepackage{color}

\usepackage{amsfonts}
\usepackage{amsmath}
\usepackage{amssymb}
\usepackage{bm}
\usepackage{dcolumn}
\usepackage{graphicx}
\usepackage[latin1]{inputenc}
\usepackage{latexsym}
\usepackage{rotating}
\usepackage{hyperref}

\newcommand\be{\begin{equation}}
\newcommand\ba{\begin{eqnarray}}
\newcommand\ee{\end{equation}}
\newcommand\ea{\end{eqnarray}}

\begin{document}

\title {Running of the Spectrum of Cosmological Perturbations in String Gas Cosmology}

\author{Robert Brandenberger}
\email{rhb@physics.mcgill.ca}
\affiliation{Physics Department, McGill University, Montreal, QC, H3A 2T8, Canada}

\author{Guilherme Franzmann}
\email{guilherme.franzmann@mail.mcgill.ca}
\affiliation{Physics Department, McGill University, Montreal, QC, H3A 2T8, Canada}

\author{Qiuyue Liang}
\email{lc9629@mail.ustc.edu.cn}
\affiliation{CAS Key Laboratory for Research in Galaxies and Cosmology, Department of Astronomy,
University of Science and Technology of China, Chinese Academy of Sciences, Hefei, Anhui 230026, China}

\date{\today}

%%%%%%%%%%%%%%%%%%%%%%%%%%%%%%%%%%%%%%%%%%%%%%%%%%%%%%%%%%%%%%%%%%%%%%%%%%%%%%%%%%%%%%%%%%%%%%
\begin{abstract}

We compute the running of the spectrum of cosmological perturbations in
String Gas Cosmology, making use of a smooth parametrization of
the transition between the early Hagedorn phase and the later
radiation phase. We find that the running has the same sign as in
simple models of single scalar field inflation. Its magnitude is
proportional to $(1 - n_s)$ ($n_s$ being the slope index of the spectrum), 
and it is thus parametrically larger than for inflationary cosmology, where
it is proportional to $(1 - n_s)^2$.

\end{abstract}
%%%%%%%%%%%%%%%%%%%%%%%%%%%%%%%%%%%%%%%%%%%%%%%%%%%%%%%%%%%%%%%%%%%%%%%%%%%%%%%%%%%%%%%%%%%%%%

\maketitle

%%%%%%%%%%%%%%%%%%%%%%%%%%%%%%%%%%%%%%%%%%%%%%%%%%%%%%%%%%%%%%%%%%%%%%%%%%%%%%%%%%%%%%%%%%%%%%
\section{Introduction} 

String Gas Cosmology (SGC) \cite{BV} is a scenario of early universe cosmology based
on fundamental principles of superstring theory (for reviews see e.g. \cite{SGCrevs}). 
It is assumed that the early universe is a
hot gas of fundamental superstrings on a spatial manifold of compact topology\footnote{Specifically,
it is assumed that the first homotopy group of the spatial sections is non-vanishing which is the
case if space is toroidal.}. In this case, it can be shown that the T-duality symmetry 
of string theory (see e.g. \cite{Pol}) implies that the temperature has the following symmetry
\be \label{Tdual}
T(R) \, = \, T\left(\frac{1}{R}\right) 
\ee
where $R$ is the radius of the spatial torus measured in units of the string length.

More generally, there is a maximal temperature for a gas of fundamental strings in
thermal equilibrium, the Hagedorn temperature $T_H$ \cite{Hagedorn}. In the presence
of the duality symmetry (\ref{Tdual}) a plot of temperature as a function of $R$ will
have the form given in Fig. 1, with a characteristic plateau where $T \sim T_H$ whose
width depends on the total entropy of the system \cite{BV}. This is the {\it Hagedorn phase}.

\begin{figure}[h]
    \centering
    \includegraphics[scale = 0.4] {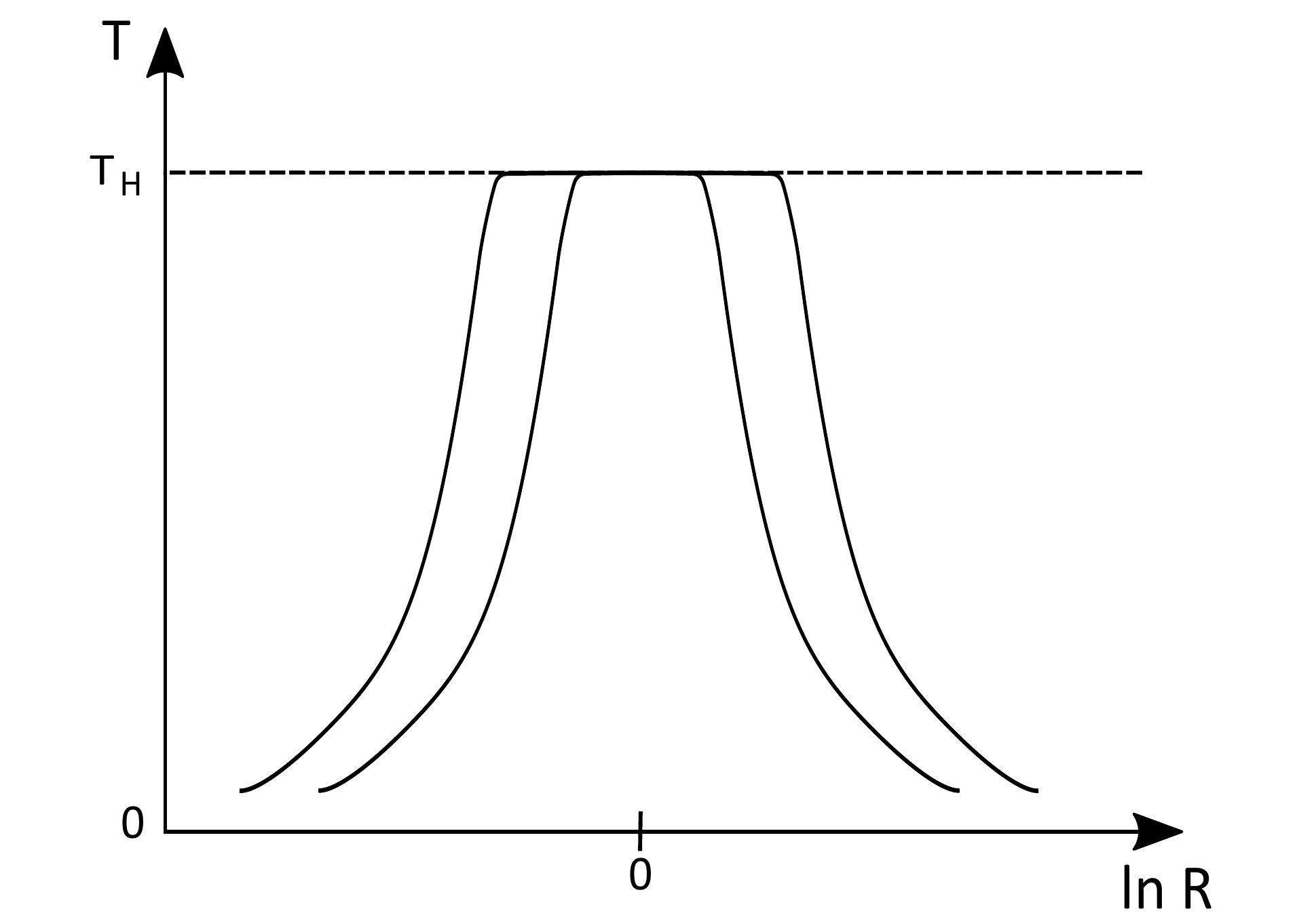}
    \caption{T versus $\log{R}$ for Heteroic superstrings. Different curves are obtained for different 
    values of the entropy - the larger the entropy the larger the plateau.}
    \label{fig1}
\end{figure}

It is natural \cite{BV} to expect the self-dual point $R = 1$ to be a metastable fixed point. If
we follow the evolution starting from a hot gas of strings, we expect the radius $R$ to
remain approximately constant for a long period of time until in three spatial dimensions
the winding modes annihilate into string loops, allowing the scale factor $a(t)$ of those
dimensions of space to expand, as illustrated in Fig. 2. 

\begin{figure}[h]
    \centering
    \includegraphics[scale = 0.5] {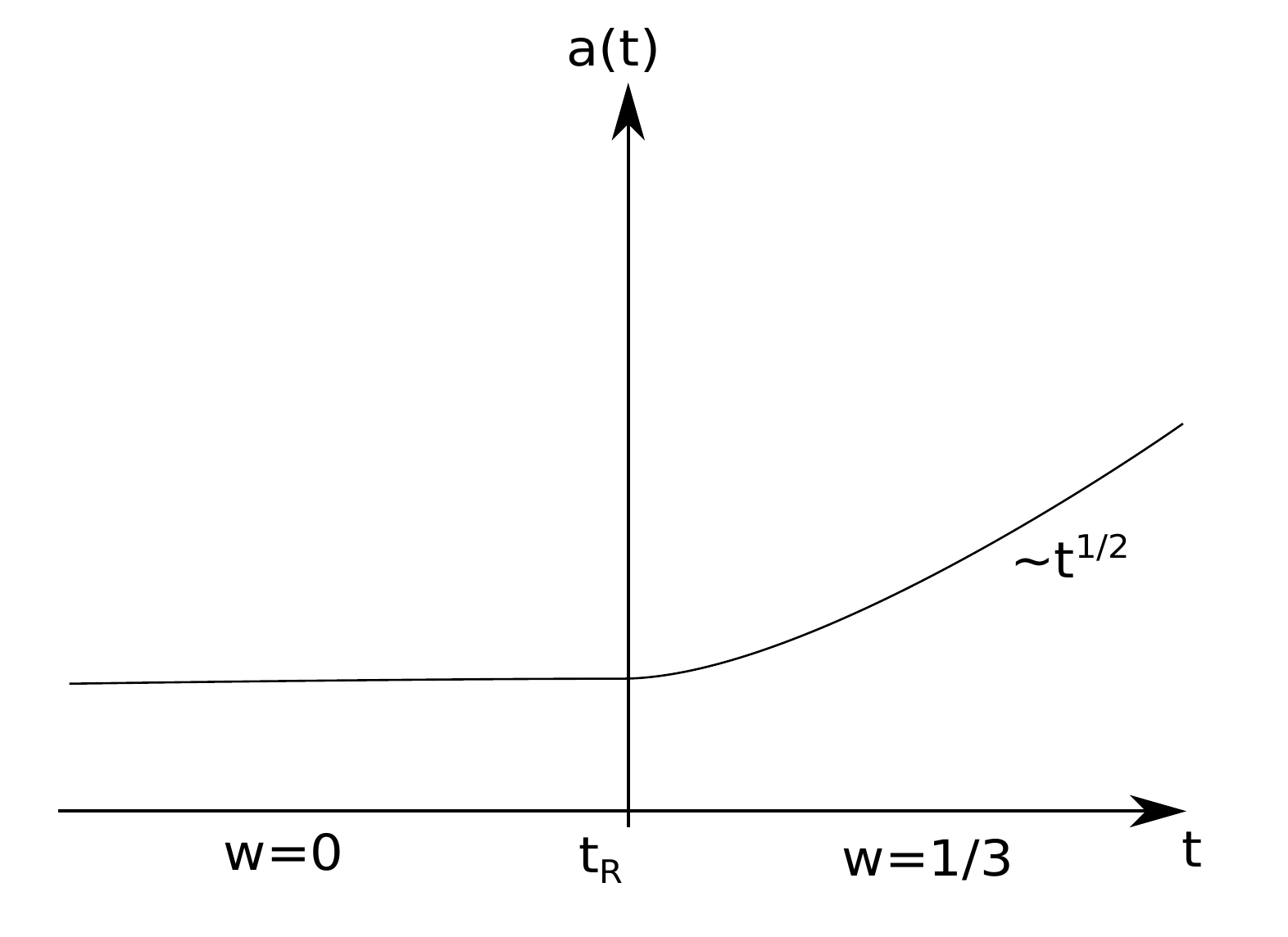}
    \caption{Sketch of the evolution of the scale factor $a(t)$ (vertical axis) as a function
    of time $t$ (horizontal axis). The time $t_R$ denotes the end of the Hagedorn phase.} 
    \label{fig2}
\end{figure}

As first discussed in \cite{NBV}, string gas cosmology leads to a theory for the origin
of structure in the universe which is different from the inflationary universe scenario.
The relevant space-time diagram is sketched in Fig. 3. Here, the vertical axis is time,
and the horizontal axis represents distance in physical coordinates. The heavy solid
curve represents the Hubble radius $H^{-1}(t)$ which is infinite in the Hagedorn
and expands as in Standard Big Bang cosmology after the phase transition is
completed. As is evident, all length scales $k$ ($k$ denoting the comoving
wavenumber) start their evolution inside the Hubble radius. They exit the Hubble 
radius at a time $t_H(k)$ which is close to the time $t_R$ but increases slightly
as $k$ increases. Since the Hagedorn phase is filled with a hot gas of fundamental
strings, the fluctuations in this phase will be thermal fluctuations of a string gas.

\begin{figure}[h]
    \centering
    \includegraphics[scale = 0.4] {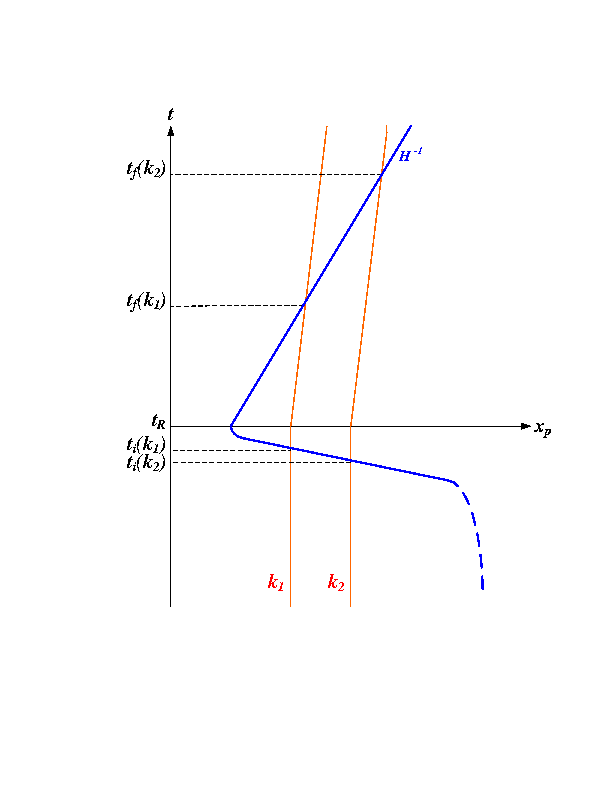}
    \caption{Space-time sketch of string gas cosmology. The vertical axis is time, with $t_R$ 
    denoting the end of the Hagedorn phase. The period $t < t_R$ corresponds to the
    quasi-static Hagedorn phase, the period $t > t_R$ to the radiation-dominated
    phase of expansion of Standard Big Bang cosmology. The horizontal axis is physical length. The
    heavy solid line represents the Hubble radius, the light solid lines which are vertical
    for $t < t_R$ correspond to the wavelengths corresponding to various comoving
    wavenumbers $k$. The wavelength equals the Hubble radius (the modes ``exit'' the
    Hubble radius) at a time $t_H(k)$ and then re-enter late in the phase of Standard
    cosmology.}
    \label{fig3}
\end{figure}

From Fig. 3 it is clear that all scales originate inside the Hubble radius. Hence,
it is possible to have a causal structure formation scenario, as in the case
of inflationary cosmology. In contrast to the case of inflation, however, the
fluctuations are thermal and not vacuum since the state of matter in the
Hagedorn phase is a hot gas of strings. As was shown in \cite{NBV} 
(see also \cite{Nayeri, BNPV1}), thermal fluctuations of strings in the Hagedorn
phase lead to an approximately scale-invariant spectrum of cosmological
perturbations. The amplitude of the spectrum on a scale $k$ is set by \cite{NBV}
$(1 - T/T_H)^{-1}$, where $T$ is the temperature when the mode $k$ exits
the Hubble radius. Since this temperature decreases as $k$ increases, a slight
red tilt of the spectrum results.

SGC also generates gravitational waves. Their amplitude is set by the
factor $(1 - T/T_H)$ \cite{BNPV2}, and hence the spectrum - while approximately
scale-invariant - has a slight blue tilt, in contrast to inflationary scenarios
where a red tilt is predicted \footnote{By invoking matter violating the
usual energy conditions it is possible to obtain a blue spectrum \cite{Galileon},
but SGC remains distinguishable from such models \cite{YiWang}.}. Thus,
the tilt of the gravitational wave spectrum is a key characteristic of SGC.

There are two free parameters in SGC: the ratio of the string length to the
Planck length, and the the value of $(1 - \frac{T}{T_H})$ during the Hagedorn
phase. These parameters can be fit to the observed amplitude and slope
of the spectrum of scalar metric fluctuations. Then, the tensor to scalar
ratio $r$ and the slope of the tensor spectrum $n_t$ are determined. One
obtains
\be \label{consistency1}
n_t \, \simeq \, (1 - n_s) \, ,
\ee
where $n_s$ is the slope of the scalar spectrum. However, the value of $r$
is predicted to be well below the current observational limits. Hence,
we are currently not able to test the consistency relation (\ref{consistency1}),
and we will not be able to in the near future. Hence, it is interesting
to search for other ways to observationally distinguish between the
predictions of SGC and those of simple single field slow-roll inflation
models. 

In this paper we study the {\it running} of the spectrum of cosmological
perturbations. The running\footnote{We use the sign convention
that the running is negative if the slope of the spectrum becomes
smaller (i.e. the spectrum becomes more red) as $k$ increases.}
has recently been shown to be a diagnostic to
differentiate between inflationary models and matter bounce cosmologies
\cite{EWE} (see e.g. \cite{BP} for a recent review of bouncing models).
Whereas the running has a negative sign in the case of inflation, it has a typical
positive sign in the case of a matter bounce model. In the present work
we compute the running of the spectrum assuming a simple smooth  
transition between the Hagedorn phase and the expanding radiation
phase of Standard Big Bang cosmology. We find that the running has
the same sign as in the case of inflation, but that it is parametrically
larger. In the case of inflation one has (see e.g. \cite{EWE})
\be \label{infl-pred}
\alpha_s \, \sim \, - (n_s - 1)^2 \, 
\ee
whereas SGC predicts
\be \label{sgc-predict}
\alpha_s \, \sim \, - (1 - n_s) \, .
\ee
This consistency relation is much easier to confront with observational
results. 

\section{Setup} 

The slope and running of the spectrum of cosmological perturbations are defined by
\ba
n_s-1 \, & \equiv & \, \frac{d \text{lnP}_\Phi(k)}{d \text{ln}k}|_{k=aH} \, = \, 
\frac{k}{P_\Phi}\frac{dP_\Phi}{dk} \label{gen-slope} \\
\alpha_s \, & \equiv & \frac{d^2 \text{ln}P_\Phi(k)}{d\text{ln}k^2}|_{k=aH} \label{gen-running} \\
& = & \,  \frac{k}{P_\Phi}\frac{dP_\Phi}{dk} -
\left(\frac{k}{P_\Phi}\frac{dP_\Phi}{dk}\right)^2 + \frac{k^2}{P_\Phi}\frac{d^2 P_\Phi}{dk^2} \, . \nonumber
\ea
In the above, $P_{\Phi}$ is the dimensionless power spectrum of the relativistic potential $\Phi({\bf x}, t)$
which in longitudinal gauge characterizes the fluctuation of the metric in the following
way (see e.g. \cite{MFB})
\be
ds^2 \, = \, - (1 + 2 \Phi)dt^2 + (1 - 2 \Phi)a^2(t)\delta_{ij}dx^idx^j \, ,
\ee
where $ds$ is the proper distance, $t$ is physical time and $x^i$ are the comoving
spatial coordinates (the index $i$ runs over the spatial indices 1,2,3 and $\delta_{ij}$
is the usual Kronecker delta symbol). The symbol $k$ stands for the comoving momentum
magnitude (we are assuming spatial isotropy), and we are evaluating the derivatives at
the Hubble scale (our pivot scale).

The current constraints (68\% confidence levels) on the slope and the running from the Planck
temperature and lensing analysis \cite{Planck} are
\ba
n_s -1 \, &=&  \, -0.032 \pm 0.006 \\
\alpha_s \, &=& \,  - 0.003 \pm 0.007 \, .
\ea
Note that in deriving these constraints it is assumed that the contribution of 
gravitational waves to the spectrum of fluctuations is negligible, which is not a
good approximation in SGC. Hence, the effective error bars are larger.

In a typical slow roll inflation, the power spectrum of the curvature
perturbation ${\mathcal{R}}$ (which up to a numerical constant equals
the potential $\Phi$ at late times) has the form:
\be
\mathcal{P}_{\mathcal{R}}(k) \, = \, \frac{H^2}{\epsilon}
\ee
where the right-hand side is evaluated when the scale $k$ exits the
Hubble radius during inflation, and $\epsilon \equiv -d lnH/d ln a$ is the slow roll parameter. 
Using the e-folding number $N$ as the time parameter, $\epsilon$ can be
parametrized as (see e.g. \cite{Mukh2})
\be
\epsilon(N) \, = \, \frac{1}{(N + 1)^p} \, ,
\ee
where $p = 1$ for chaotic inflation \cite{Linde} and $p = 2$ for plateau inflation \cite{Starob}.
The spectral index and running  then take on the following form (see e.g. \cite{Limin})
\ba
n_s -1 \, &=& \,  -2\epsilon +\frac{\epsilon,_{\mathcal{N}}}{\epsilon} \\
& & = \, -\frac{2}{(\mathcal{N}+1)^p}-\frac{p}{\mathcal{N}+1} \nonumber \\
\alpha_s \, &=& \, 2\epsilon,_{\mathcal{N}} - \frac{\epsilon,{_\mathcal{NN}}}{\epsilon}+ (\frac{\epsilon,_{N}}{\epsilon})^2 \nonumber \\
& & = \, -\frac{2p}{(\mathcal{N}+1)^{p+1}}-\frac{p}{(\mathcal{N}+1)^2} \, .
\ea
and this will then yield the quadratic relation (\ref{infl-pred}) between the
running $\alpha_s$ and $n_s -1$. A small deviation of the spectrum from
scale-invariance will lead to a very small running of the spectrum.

As we show in the following section, the result in string gas cosmology is different.

\section{Analysis}

In string gas cosmology the fluctuations are of thermal origin. As derived in \cite{NBV},
the spectrum of cosmological perturbations on a scale $k$ is given by
\be \label{sgc-spectrum}
P_{\Phi}(k) \, = \, 16\pi^2G^2 \frac{T(k)}{l_s^3} \frac{1}{1-T(k)/T_H} \, \approx \, C \frac{1}{\epsilon(k)}
\ee
where $C$ is a constant and we have defined 
\be
\epsilon(k) \, \equiv \, \left(1 - \frac{T(k)}{T_H}\right) \, .
\ee
Here, $T(k)$ is the temperature of the string gas when the scale $k$ exits the Hubble radius
at the end of the Hagedorn phase (see Fig. 3).

Since all scales exit the Hubble radius at temperatures close to the Hagedorn
temperature, the overall shape of the spectrum is scale-invariant. Since modes
with larger values of $k$ exit slightly later, when the temperature is slightly lower,
the value of $\epsilon$ is slightly larger, and thus a slight red tilt of the spectrum
results, as in inflationary cosmology. Here, we wish to compute the running of
the spectrum.

Quantitatively, from (\ref{sgc-spectrum}) and from the definitions of the slope and
running of the spectrum we obtain
\ba
n_s - 1 \, &\simeq& \, - \frac{\epsilon^{\prime}}{\epsilon} \label{slope} \\
\alpha_s \, &\simeq& \, - \frac{\epsilon^{\prime \prime} \epsilon - (\epsilon^{\prime})^2}{\epsilon^2} 
\label{running} \, ,
\ea
where the prime indicates a derivative with respect to ${\rm ln}k$. In the above,
the value of $\epsilon(k)$ is determined by $\epsilon(T_H(k))$, where $T_H(k)$ is
the temperature when the mode $k$ exits the Hubble radius.

Unfortunately the dynamics of space-time during the Hagedorn phase is not
yet known. Hence, we need to resort to a parametrization of the dynamics, based
on the thermodynamics of a gas of strings represented in Fig. 1, and the
evolution of the scale factor assumed (see Fig. 2).

In the Hagedorn phase, the radius of our universe is constant, while in the 
radiation-dominated phase, the scale factor increases with time like $R(t)\sim t^{1/2}$. 
We first use a quadratic polynomial to connect the two phases. We take
the time $t = 0$ to be the end of the Hagedorn phase, and $t_0$ to be
the end of the transition phase:
\begin{equation}
\begin{aligned}
R(t) = \left\{
\begin{aligned}
& c,& t\leqslant 0\\
& c +\alpha \left(\frac{t}{t_0}\right)^2,& 0<t\leqslant t_0\\
& c +\beta \left(\frac{t}{t_0}+\gamma\right)^{1/2},& t>t_0  
\end{aligned}
\right.
\end{aligned}
\end{equation}
where $\beta =2\alpha$ to make the function continuous. In addition, we can 
choose $\gamma=-3/4$ to render the first derivative continuous at $t_0$ .

In the expressions (\ref{slope}) and (\ref{running}) for the slope and the running
of the spectrum, $\epsilon(k)$ is evaluted at the temperature when the mode $k$
exits the Hubble radius. We can thus relate the derivative with respect to $k$
with a derivative with respect to radius $R$ (the scale factor). We can then
use the Hubble radius crossing condition
\be
k R^{-1} \, = \, \frac{{\dot R}}{R}
\ee
and the expression for $R(t)$ to relate the derivative of $R$ with respect to
$k$ by the derivative of $t_H(k)$ with respect to $k$, namely
\be
\frac{\partial R}{\partial k} \, = \, k \frac{\partial t}{\partial k}
\ee
where $t = t_H(k)$. Combining these results we then obtain
\ba
n_s -1 \, &\approx& \,  -\frac{d\text{ln}\epsilon}{d\text{ln}k} =- \frac{k}{\epsilon} \frac{d\epsilon}{d R}_{|R_0}\frac{dR}{dk} \, = \, - \frac{k^2}{\epsilon} \beta_1\frac{dt}{dk} \\
\alpha_s \, &\approx& \, -\frac{d^2\text{ln}\epsilon}{d\text{ln}k^2} \\
&=& \, (n_s -1)\left(2+k\frac{d^2 t }{dk^2}\right)+ (n_s -1)^2\left(1-\frac{\epsilon \beta_2}{\beta_1 ^2}\right) \nonumber
\ea
where $\beta_1$ and $\beta_2$ are first and second derivative of $\epsilon(R)$,
evaluated at the pivot scale $R_0$, and they are simply related to the first and second time
derivatives of the temperature function $T(R)$ after multiplying by $- T_H$. Large
length scales such as those which are probed in cosmological observations exit
the Hubble radius early, when $T(k)$ is close to the Hagedorn temperature. From
the shape of $T(R)$ in Fig. 1 it then follows that both $\beta_1$ and $\beta_2$ are 
positive numbers. Making use of
\be
t_H(k) \, = \, \frac{t_0^2}{2\alpha}k 
\ee
we finally get
\begin{equation}
\begin{aligned}
&n_s -1 = - \frac{k^2}{\epsilon}\frac{t_0^2}{2\alpha}\beta_1\\
&\alpha_s = -\frac{k^2}{\epsilon}\frac{t_0^2}{2\alpha}2\beta_1 + \left(\frac{k^2}{\epsilon}\frac{t_0^2}{2\alpha}\beta_1\right)^2\left(1-\frac{\epsilon \beta_2}{\beta_1 ^2}\right) \, .\\
\end{aligned}
\end{equation}

Since $\beta_1$ and $\beta_2$ are both positive, we get a negative running. and, Since 
$n_s -1 \approx - 0.03$ from the most recent CMB data, the $(n_s -1)^2$ terms are
negligible. Hence we predict a linear relation between $n_s -1$ and $\alpha_s$ 
as in (\ref{sgc-predict}).

We also considered a more general parametrization of $R(t)$ in which R is proportional 
to $(t/t_0)^n$, and we get a similar linear relation
\begin{equation}
\alpha_s \, = \, (n_s-1)\left[2-\frac{n-2}{n-1}+(n_s - 1)\left(1-\frac{\epsilon \beta_2}{\beta_1^2}\right)\right]
\end{equation}
between the running and the tilt of the spectrum. Different models just influence 
the $\mathcal{O}(1)$ coefficient in the linear relation between running and spectral index.
Note that if $n$ is negative, a blue spectrum of cosmological perturbations results.

\section{Conclusions and Discussion}

In conclusion, we have calculated the running of the power spectrum in string gas cosmology (SGC)
and obtain a negative running. In addition, our parametrizations of $R(t)$ and $T(R)$
yield a linear relationship between the spectral index and running. This string gas
cosmology consistency relation is already in the realm of what can be tested with
current data. In fact, it may appear that our predictions are already in tension with the
latest Planck results. However, in string gas cosmology the contribution of gravitational
waves is not negligible, and hence the limits of \cite{Planck} are not directly applicable.
It would be interesting to do an analysis of the current data allowing amplitude,
slope, running and gravitational wave contributions to all vary as predicted in SGC.

While we have presented our results for specific parametrizations of $R(t)$
and $T(t)$, our main result - namely the linear dependence of the running of the
spectrum on the deviation of the spectrum from scale-invariance - is quite
general. We have explored an extended space of parametrizations.
Note that since we care most about the large scale perturbations, 
we focus on $t\sim 0$. In particular, it can be shown that for an arbitrary 
interpolation function between the two phases we still obtain a linear relation 
between the running and $n_s - 1$, having different interpolation functions 
influencing only the $\mathcal{O}(1)$ coefficient.

Given we are predicting
a linear relation between running and slope of the spectrum of
cosmological perturbations,  
let us step back and ask
why one obtains a quadratic relation (\ref{infl-pred}) in simple inflationary models:
Take a back at the equation (\ref{gen-running}), we find that 
$ \frac{d^2P_\Phi}{dk^2} k^2/P_\Phi$ is positive in inflation and cancels against the first term, 
leaving the second term which is approximately proportional to $|n_s -1|^2$.
But in string gas model, both the first term and the third term are negative, and hence
there is no cancellation.  Considering that the second term is proportional to 
$|n_s-1|^2$ it will be negligible compared to the first term. However, to make a quick transition, 
$\beta_2$ should be a large amount, thus we cannot neglect the third term. 
If we suppose that $\beta_2$ is small, corresponding to a smooth and slow transition, 
this will give us the prediction $|\alpha_s| \sim |n_s -1|$.

\section*{Acknowledgement}
\noindent

The research at McGill is supported in
part by funds from NSERC and from the Canada Research Chair program. 
GF acknowledges financial support from CNPq (Science Without Borders). 
QL acknowledges financial support from The University of Science and Technology of China.

\end{document}